\documentclass[twocolumn,showpacs,superscriptaddress,twoside,prl]{revtex4}
\usepackage{mathrsfs, doi, hyperref}
\usepackage{amssymb, amsbsy, amsmath, latexsym, dsfont, array, layout, graphicx,mathrsfs,braket,amsfonts,amsthm, amssymb,graphicx,subfigure, natbib, youngtab,color,xcolor}

\newcommand\rket[1]{\vert\, #1 )}
\newcommand\rbra[1]{( #1 \vert}

\hypersetup{
    colorlinks=true,       
    linkcolor=blue,          
    citecolor=blue,        
    filecolor=blue,      
    urlcolor=blue           
}

\begin{document}

\title{SU(3) Quantum Interferometry with Single-Photon Input Pulses}
\author{Si-Hui Tan}
\email{sihui.tan@gmail.com}
\affiliation{Data Storage Institute, 5 Engineering Drive 1, Singapore, Singapore 117608}
\author{Yvonne Y. Gao}
\affiliation{Department of Physics and Applied Physics, Yale University, New Haven, Connecticut 06520, USA}
\author{Hubert de Guise}
\affiliation{Department of Physics, Lakehead University, Thunder Bay, Ontario P7B 5E1, Canada}
\author{Barry C. Sanders}
\affiliation{Institute for Quantum Information Science, University of Calgary, Alberta, Canada T2N 1N4}

\begin{abstract}
We develop a framework for solving the action of a three-channel passive optical interferometer on single-photon pulse inputs to each channel using SU(3) group-theoretic methods,
which can be readily generalized to higher-order photon-coincidence experiments.
We show that features of the coincidence plots vs relative time delays of photons yield information
about permanents, immanants, and determinants of the interferometer SU(3) matrix.
\end{abstract}

\date{\today}
\maketitle

An optical interferometer is a coherent scatterer of light comprising passive optical elements such as phase shifters, mirrors, and beam splitters.
The interferometer operates in the quantum regime by injecting nonclassical light 
into the input ports and yields nonclassical light from the output ports.
The most famous quantum optical effect is 
the two-photon Hong-Ou-Mandel dip~\cite{HOM87},
with the ``dip'' corresponding to extinction of photon coincidences 
at the two output ports of a balanced beam splitter given two identical input photons
at each input port.
Instead, the two photons exit in a superposition state of leaving together from each of the two output ports.
Important applications include characterizing distinguishability between pairs of independent photons~\cite{Man99},
measuring coherence~\cite{HBT+08} and purity \cite{CLS10} of single photons,
producing two-photon entanglement~\cite{Gio06},
performing dense coding~\cite{MWKZ96} and single-qubit quantum fingerprinting~\cite{HBM+05},
and creating nondeterministic nonlinear gates in optical quantum computing~\cite{KLM01}.

Generalizing to higher-order photon-coincidence dips will be valuable in many
ways including 
determining distinguishability between multiple photons simultaneously,
computing the permanent of special unitary (SU) matrices through photon
coincidence dips~\cite{BSM+04},
and sampling permanents of submatrices of SU matrices to demonstrate that boson sampling is probably hard for a classical computer \cite{AA11}.
Experimentally eight individual photons have been manipulated with exquisite control~\cite{YWX+12}, temporal distinguishability of four- and six-photon states have been characterized~\cite{XHSZOG+06}, 
and three photons in two and three coupled interferometers have shown
nonclassical interference~\cite{SRZ+06,MTS+12},
so higher-order photon coincidence dips are feasible well beyond the two-photon two-channel case.
Although coherent scattering of Fock states has been analyzed theoretically 
for various instances of interferometers~\cite{Cam00,LB05} and even a zero-transmission condition for $2N$-port devices \cite{TTMMB+2010},
a full multimode analysis of multichannel interferometry incorporating realistic source and detector spectral responses \cite{Ou08}, but allowing for arbitrary configurations has not yet been fully studied
despite its necessity for quantitative studies of these nonclassical interferometric systems.

Here 
we highlight the rich group theoretical structure of photon interferometry and its direct relationship to permutation symmetries and associated matrix functions
by analyzing
the problem
of photon coincidence probabilities at the output of
any passive three-channel interferometer
[whose action is represented by the SU(3) matrix]
acting on single-photon pulse inputs. Such an approach was pioneered by Campos {\it et al.~}in a four-port result \cite{CSBT89} for SU(2) interferometers. 
Our approach creates a pathway for discussing 
matrix functions in arbitrary $n$-mode interferometers beyond the case of three modes reported here. Our goal is to to develop a full realistic theory for photon coincidences that necessarily accommodates multimode photon pulses, multimode detection and photon delays.

Input photons can reach the detectors by various paths. If the amplitudes and phases of these paths interfere destructively, a coincidence dip occurs; if all paths exactly cancel out as a result of suitably chosen amplitudes and phases, a complete dip occurs. The contribution to the coincidence rate of photons with different frequencies is related by permutation symmetries. These  contributions enter into our expressions as weighted contributions of the immanants, including the permanent as a special case, to the coincidence rate.


A single-photon pulse with light-source spectral function
$\tilde{\phi}^S(\omega)$
in one spatial mode,
or channel, is
\begin{equation}
\label{eq:1photon}
	\ket{1}^S=\int d\omega \tilde{\phi}^S(\omega)\rket{1(\omega)},\;
		\int d\omega |\tilde{\phi}^S(\omega)|^2=1
\end{equation}
for $\rket{1(\omega)}\equiv\hat{a}^\dag(\omega)\ket{0}$
with $\hat{a}(\omega)$ the creation operator at frequency~$\omega$ satisfying
$[\hat{a}_k(\omega_i),\hat{a}_l^\dagger(\omega_j)]
	=\delta_{kl}\delta(\omega_i-\omega_j)\openone$,
$k$ and~$l$ labels for distinct channels,
and~$\openone$ the identity operator.
The parenthetical (rounded) bra-ket notation distinguishes frequency-explicit states
from other states.
For convenience, we choose a Gaussian spectral function
\newpage
\begin{equation}
\label{eq:gaussian}
	\tilde{\phi}(\omega)
		=\left(2\pi \sigma_0^2\right)^{-1/4}
\exp[-(\omega-\omega_0)^2/ (4\sigma_0^2)],
\end{equation}
but our approach accommodates any 
spectral function.

The detector has a response function~$\tilde{\phi}^D(\omega)$,
possibly different for each output port.
For convenience we take~$\tilde{\phi}^{D}(\omega)$ to be Gaussian,
similar to~(\ref{eq:gaussian}),
but perhaps with different mean and variance;
the variance is inversely related to the detector integration time.
For
\begin{equation}
\label{eq:nD}
	\ket{n}
		:=\frac{1}{\sqrt{n!}}\left(\prod^n_{j=1}\int d\omega_j
			\tilde{\phi}^D(\omega_{j})\hat{a}^\dagger(\omega_{j})\right)\ket{0}
\end{equation}
an $n$-photon state corresponding to a superposition of single-photon states in 
different infinitesimal frequency modes,
the ideal detector executes projective measurements of the type
$\Pi_n:=|n\rangle\langle n|$, $\sum_{n=0}^\infty\Pi_n=\openone$.
Detector inefficiency arises due to a spectral mismatch between source and detector spectral functions: the probability of detecting a single source photon is
$|\int d\omega[\tilde{\phi}^S(\omega)]^*\tilde{\phi}^D(\omega)|^2 \leq 1$.

We assume identical photons from the source, 
with distinguishability introduced by time delays of the pulses.
A tunable time delay  of~$\tau$ on the single-photon pulse
is expressed in the Fourier domain by a phase shift
$\ket{1}\mapsto\int d\omega\tilde{\phi}(\omega)\exp[-i\omega\tau]\rket{1(\omega)}$,
and the inverse Fourier transform of
$\tilde{\phi}(\omega)\text{exp}[-i\omega\tau]$
is~$\phi(t-\tau)$.

Now that we have established the single-mode source and detector formalism,
we consider two photons impinging on the two input ports of the passive two-channel SU(2) interferometer resulting in a photon-coincidence dip.
This transformation can be expressed as
\begin{align}
\label{su2}
	R(\Omega)
		=&\begin{pmatrix}
			e^{-i(\alpha+\gamma)}\cos\frac{\beta}{2}
				&-e^{-i(\alpha-\gamma)}\sin\frac{\beta}{2}\\
			e^{i(\alpha-\gamma)}\sin\frac{\beta}{2}
				& e^{i(\alpha+\gamma)}\cos\frac{\beta}{2}
			\end{pmatrix}
\end{align}
with~$\Omega$ comprising the three Euler angles
$\alpha,\beta,\gamma$,
and the optical-element transformations are assumed to be independent of frequency.

The input state~$\ket{11}^S$ has one photon in each port, 
i.e., a tensor product of two single-photon source states~(\ref{eq:1photon}) with time delay $\tau$ between modes 1 and 2.
As we need to convert the direct product of irreps for single-photon interferometer inputs
to a direct-sum decomposition, Young diagrams are valuable, especially for SU($n$) interferometry
with $n>2$ as these cases can be complicated to construct.
For a system with $p$ photons, Young 
diagrams comprise $p$ $\square$s arranged in columns of most $n$ boxes that
represent various regular partitions of~$p$.  These partitions \emph{simultaneously} label representations of SU($n$) \emph{and} of the
permutation group of $p$ objects.

The representation of any SU(2) matrix on two photons with frequencies $\omega_1$ and $\omega_2$ decomposes as
\begin{equation}
	{\renewcommand{\arraystretch}{0.8}
	{\renewcommand{\arraycolsep}{0.825pt}
	\begin{array}{ccccccc}
	\Yboxdim{6pt}\yng(1)&\otimes &\Yboxdim{6pt}\yng(1) &\to &
	\Yboxdim{6pt}\yng(2)&\oplus  & \Yboxdim{6pt}\yng(1,1)\\
			\\
	(1)                          &\otimes & (1)                        & \to & 
	(2)                         & \oplus &(0) 
\end{array}}}
\label{eq:SU2decomposition}
\end{equation}
with the second row giving the standard labels of the irreps in terms of 
highest weight.
Equation (\ref{eq:SU2decomposition}) is also the decomposition into a symmetric triplet and antisymmetric singlet $3\oplus 1$ 
corresponding to states with angular momentum $\ell=1,0$ respectively,
and $m=-\ell,\dots,\ell$.

The two-photon state is a combination of $\ket{\ell m}$ states:
$\rket{1(\omega_1)1(\omega_2)}=\textstyle\frac{1}{\sqrt{2}}\left(\ket{00}+\ket{10}\right)$.
The single-photon state 
$\hat{a}_1^\dagger(\omega_i)\ket{0}$ transforms under SU(2) 
as $\ket{\ell=\frac{1}{2}\,m=\frac{1}{2}}_i$
and $\hat{a}_2^\dagger(\omega_i)\ket{0}$
transforms as $\ket{-\frac{1}{2}\,-\frac{1}{2}}_i$:
\begin{align}\nonumber
	R(\Omega)\hat{a}_1^\dagger
	(\omega_1)\ket{0} =&[\hat{a}_1^\dagger(\omega_1)D^{1/2}_{1/2,1/2}(\Omega) \\ &
			+\hat{a}_2^\dagger(\omega_1)
			D^{1/2}_{-1/2,1/2}(\Omega)]\ket{0}
				\nonumber\\					
	R(\Omega)\hat{a}_2^\dagger(\omega_2)\ket{0}\nonumber 
	=&[\hat{a}_1^\dagger(\omega_2)D^{1/2}_{1/2,-1/2}(\Omega) \\
	& + {\hat a_2^\dagger}(\omega_2)
	D^{1/2}_{-1/2,-1/2}(\Omega)]\ket{0}
\end{align}
with $D^\ell_{m'm}(\Omega):=\braket{\ell m'|R(\Omega)| \ell m}$ an element of 
the $(2\ell+1)$-dimensional Wigner $D$ matrix.
The two-photon transformation is thus
\begin{equation}
	R(\Omega)\rket{1(\omega_1)1(\omega_2)}
		=\textstyle\frac{1}{\sqrt{2}}\left(R(\Omega)\ket{00}+R(\Omega)\ket{10}\right).
\end{equation}

Extinguishing coincidences arising from the triplet $|10\rangle$ contribution,
which is symmetric under an $\omega_1\leftrightarrow\omega_2$ exchange,
requires that
$\bra{10}R(\Omega)\ket{11}
		=\textstyle\frac{1}{\sqrt{2}}D^{1}_{00}(\Omega)=\cos\beta=0$.
Therefore, $\beta={\textstyle{1\over2}}\pi$ and
is independent of $\alpha,\gamma$.
These parameters correspond to a balanced beam splitter 
with a relative phase shift between modes~$1$ and~$2$.
For $\alpha=\pi/4=-\gamma$,
the beam splitter is balanced and symmetric:
$B:=R(\Omega)$ for this important case.

Typically detectors at different output ports are dissimilar.
If the source pulse has carrier frequency~$\omega_0$ and bandwidth~$\sigma_0$,
and detectors have carrier frequencies~$\omega_i$ and bandwidths~$\sigma_i$
for each output mode $i$,
we introduce weighted variances
$\overline{\sigma^2}_i:=\sigma_0^2+\sigma_i^2$,
\begin{equation*}
	\overline{\varsigma^2}_i
		:= \frac{\sigma_i^2\omega_0+\sigma_0^2\omega_i}{\sigma_0^2+\sigma_i^2},\,
	\widetilde{\sigma^2}_i
		:=\left(\frac{1}{\sigma_0^2}+\frac{1}{\sigma_i^2}\right)^{-1},
\end{equation*}
and Gaussian spectral mismatch function
\vspace{-0.2cm}
\begin{equation}
	\Lambda_i
		\equiv \sqrt{\frac{2\widetilde{\sigma^2}_i}{\overline{\sigma^2}_i}}
			\exp\left(-\frac{(\omega_0-\omega_i)^2}{2\overline{\sigma^2}_i}\right).
\end{equation}
The two-mode output coincidence rate is
\begin{align}
	P_{11}
		=&^S\!\braket{11|B^\dagger\Pi_1\otimes\Pi_1B|11}^S\\
		=&\Lambda_1\Lambda_2\left| e^{-\tau^2/\widetilde{\sigma^2}_1}
			e^{i\tau\overline{\varsigma^2}_1}
				-e^{-\tau^2\widetilde{\sigma^2}_2} e^{i\tau
\overline{\varsigma^2}_2}\right|^2,
\end{align}

$P_{11}=0$ as expected for indistinguishable photons and identical detectors.

Now we proceed to three-channel 
SU(3) interferometry transformation, with the eight-parameter generalized Euler angle~$\Omega$
(assumed to be frequency independent) and factorization~\cite{KdG10}
\begin{align}\label{su3}
	R(\Omega)
		\equiv & R_{23}(\alpha_1,\beta_1,-\alpha_1)R_{12}(\alpha_{2},\beta_2,-\alpha_2)
						\nonumber\\
		&\times R_{23}(\alpha_3,\beta_3,-\alpha_3) e^{-i\gamma_1  h_1}e^{-i\gamma_2  h_2},
						\\
	{h}_1=&2n_1 -n_2-n_3,\;
	{h}_2={\textstyle{1\over2}} (n_2-n_3),\,
	\nonumber
\end{align}
with  SU(2) subgroup matrices
\begin{align}
	R_{23}(\alpha,\beta,-\alpha)
		=&\begin{pmatrix}
			1&0&0\\
			0&\cos\frac{\beta}{2}&-e^{-i\alpha}\sin\frac{\beta}{2}\\
0 & e^{i\alpha}\sin\frac{\beta}{2}&\cos\frac{\beta}{2}
		\end{pmatrix},\\
R_{12}(\alpha,\beta,-\alpha)=&\begin{pmatrix}
	\cos\frac{\beta}{2} & -e^{-i\alpha}\sin\frac{\beta}{2} & 0\\
	e^{i\alpha}\sin\frac{\beta}{2}&\cos\frac{\beta}{2}&0\\
0&0&1
		\end{pmatrix}.
\end{align}
Experimentally, the SU(2) transformation are interpreted as the sequence phase shifter-beamsplitter-phase shifter with parameters defined by the Euler angles \cite{YMK86}.

We employ Young diagram methods to determine output states given the three-photon input state
\begin{align}
\label{eq:111}
	\ket{111}^S
		=&\int d\omega_1\int d\omega_2 \int d\omega_3 \tilde{\phi}(\omega_1)\tilde{\phi}(\omega_2)\tilde{\phi}(\omega_3)\nonumber\\
		&\times e^{i\omega_2 \tau_1} e^{i\omega_3 \tau_2}\rket{1 (\omega_1)1(\omega_2)1(\omega_3)}^S
\end{align}
with time delays $\tau_1$ and $\tau_2$ between modes~$1$ and~$2$
and between modes~$1$ and~$3$, respectively.
The three-photon input state~(\ref{eq:111}) is
\begin{equation}
	{\renewcommand{\arraystretch}{0.8}
	{\renewcommand{\arraycolsep}{0.825pt}
	\begin{array}{ccccccccccccc}
	\Yboxdim{6pt}\yng(1)&\otimes &\Yboxdim{6pt}\yng(1) &\otimes &\Yboxdim{6pt}\yng(1)&\to &
	\Yboxdim{6pt}\yng(3)&\oplus  &\Yboxdim{6pt}\yng(2,1) & \oplus 
		&\Yboxdim{6pt}\yng(2,1) & \oplus & \Yboxdim{6pt}\yng(1,1,1)\\
			\\
	(1,0)                          &\otimes & (1,0)                        & \otimes& (1,0)                        &\to & 
	(3,0)                         & \oplus   & (1,1)                        & \oplus & (1,1)& \oplus &(0,0) 
\end{array}}}
\label{eq:SU3decomposition}
\end{equation}
with the second row giving the standard labels in terms of SU(3) highest weight labels $(\lambda,\mu)$.  For a diagram with $a_i$ boxes on row $i$, we have 
$\lambda=a_1-a_2$, $\mu=a_2-a_3$.

Basis states for the irrep $(\lambda,\mu)$ are denoted $\ket{(\lambda,\mu)\nu_1\nu_2\nu_3;I}$
with~$\nu_i$ the number of photons in port~$i$
and $\nu_1+\nu_2+\nu_3=\lambda+2\mu$.  The index $I$
distinguishes states with the same weight
$(\nu_1-\nu_2,\nu_2-\nu_3)$ belonging to different irreps
of the SU$_{23}(2)$ subgroup of SU$(3)$. In this notation, the three-photon state decomposes into
\begin{align}
	\big|1(\omega_1)&1(\omega_2)1(\omega_3)\big)
	= \textstyle\frac{1}{\sqrt{6}}\ket{(00)111;0}+\textstyle\frac{1}{\sqrt{6}}\ket{(30)111;1}
			\nonumber\\
&		+\textstyle\frac{1}{2}\ket{(11)111;0}_1+\textstyle\frac{1}{\sqrt{12}}\ket{(11)111;1}_1 \nonumber\\
& -\textstyle\frac{1}{\sqrt{12}}\ket{(11)111;0}_2
+\textstyle\frac{1}{2}\ket{(11)111;1}_2.
\label{inputstate}
\end{align}
The SU$(3)$ irrep $(1,1)$ occurs twice in Eq.~(\ref{eq:SU3decomposition});
thus, an additional subscript
($1$ and~$2$)
is needed to denote basis states that belong to these distinct copies of $(1,1)$. 
This input transforms
as $R(\Omega)\rket{1(\omega_1)1(\omega_2)1(\omega_3)}$ and can be expanded in terms of the appropriate SU(3) $D$ functions.

The Young diagrams also label the representations $\Yboxdim{4pt}\yng(3)$, $\Yboxdim{4pt}\yng(2,1)$, and $\Yboxdim{4pt}\yng(1,1,1)$
of $S_3$ (the six-element permutation group of three objects).
Characters of $\Yboxdim{4pt}\yng(3)$, $\Yboxdim{4pt}\yng(2,1)$, and $\Yboxdim{4pt}\yng(1,1,1)$ are needed to construct 
the permanent, immanant and determinant of a $3\times 3$ matrix~\cite{Lit50}, respectively.
Whereas the connection between coincidences at the interferometer output,
the $D$ functions,
and the permanents is clear~\cite{S04,BSM+04,AA11},
it behooves us to find a relationship between the immanants of $R(\Omega)$
and the $D$ functions of irreps of  SU(3). Using Young diagrams to denote the corresponding functions of the matrix $R(\Omega)$
constructed with frequencies of the output modes fixed with respect to those of the input modes, we observe the following.

{\it Observations 1.}--For any $R(\Omega)\in$SU(3),
\begin{align*}
	\text{Per}(R(\Omega)) \equiv \ &\Yboxdim{4pt}\yng(3)(\Omega) =D^{(3,0)}_{(111)1;(111)1}(\Omega)\\
	\text{Imm}(R(\Omega)) \equiv \ &\Yboxdim{4pt}\yng(2,1)(\Omega) =D^{(1,1)}_{(111)1;(111)1}(\Omega)+D^{(1,1)}_{(111)0;(111)0}(\Omega)\\
	\text{Det}(R(\Omega)) \equiv \ &\Yboxdim{4pt}\yng(1,1,1)(\Omega) =D^{(0,0)}_{(111)0;(111)0}(\Omega).
\end{align*}
so the full permanent of the general $3\times 3$ SU(3) matrix equals the $D$ function for irrep (3,0) with weight~$0$ input and output.
The immanant of $R(\Omega)$ is the sum of $D$ functions for the irrep (1,1) with weight~$0$ input and output, and the determinant of $R(\Omega)$ is
just the $D$ function for irrep (0,0).

{\it Observation 2.}--
For $R(\Omega)_{kj}$ the $2\times 2$ submatrix of $R(\Omega)$ 
with the $k$th row and $j$th column removed,
\begin{equation*}
	\text{Per}(R(\Omega)_{kj})
		=\Yboxdim{6pt}\yng(2)_{kj}(\Omega)= D^{(2,0)}_{(110_k)I_k;(110_j,)I_j}(\Omega)
\end{equation*}
for~$D^{(2,0)}_{(110_k)I_k;(110_j,)I_j}(\Omega)$
the $D$ function with $0$ in output state $k$ and $1$ otherwise,
and $0$ in the input state $j$ and $1$ otherwise.
For instance,
\begin{align}
	\text{Per}(R(\Omega)_{13})=&\Yboxdim{6pt}\yng(2)_{13}(\Omega)=D^{(2,0)}_{(011)1;(110)(\frac{1}{2})}(\Omega).
\end{align}

In view of observation 1, we consider measuring coincidences for the 
monochromatic (continuous-wave) three-photon input state $\rbra{1(\omega_i)1(\omega_j)1(\omega_k)}^S$.  For the special case of
$(\omega_i,\omega_j,\omega_k)=(\omega_1,\omega_2,\omega_3)$, we find
\begin{align}
	^S\big (1(\omega_1)&1(\omega_2)1(\omega_3)\big|R(\Omega)
		\rket{1(\omega_1)1(\omega_2)1(\omega_3)}^S
				\nonumber \\
	=&\textstyle\frac{1}{6}D^{(3,0)}_{(111)1;(111)1}(\Omega)
		+\textstyle\frac{1}{6}D^{(0,0)}_{(111)0;(111)0}(\Omega)
				\nonumber \\
	&+\textstyle\frac{1}{3}(D^{(1,1)}_{(111)1;(111)1}(\Omega)
		+D^{(1,1)}_{(111)0;(111)0}(\Omega))
				\nonumber\\ \label{eq:weightedsum}
	=&\textstyle\frac{1}{6}\Yboxdim{4pt}\yng(3)(\Omega)
		+\textstyle\frac{1}{3}\Yboxdim{4pt}\yng(2,1)(\Omega)
		+\textstyle\frac{1}{6}\Yboxdim{4pt}\yng(1,1,1)(\Omega).  
\end{align}
As Eq.~(\ref{eq:weightedsum}) is covariant under permutation of the output frequencies,
detecting photons of frequencies 
$(\omega_i,\omega_j,\omega_k)$
in output ports $(1,2,3)$, respectively, implies that
\begin{align}
	D^{(\lambda,\mu)}_{(111)J,(111)I}(\Omega)\equiv & \bra{(\lambda,\mu)(111)J}R
(\Omega)\ket{(\lambda,\mu)(111)I}\nonumber \\
	\quad \mapsto& \bra{(\lambda,\mu)(111)J} ({\wp_{ijk}})^{-1}\\
	& \times R(\Omega)\ket{(\lambda,\mu)(111)I}, 
\end{align}
with $\wp_{ijk}$ the permutation operator with action
$\wp_{ijk}(\omega_1,\omega_2,\omega_3)\mapsto(\omega_i,\omega_j,\omega_k)$;
the action of $\wp_{ijk}$ on
SU(3) basis states is known from ~\cite{RSdG99}. 
Permuting output frequencies affects permutations of the rows in the SU(3) matrix.
The permanent thus remains unchanged,
the determinant picks up a sign for odd permutations,
and the immanant transforms in a more complicated way according to the two-dimensional irrep of $S_3$.
Equation~(\ref{eq:weightedsum}) remains valid provided the above changes are implemented.

\begin{figure}
        \includegraphics[width=4.2cm]{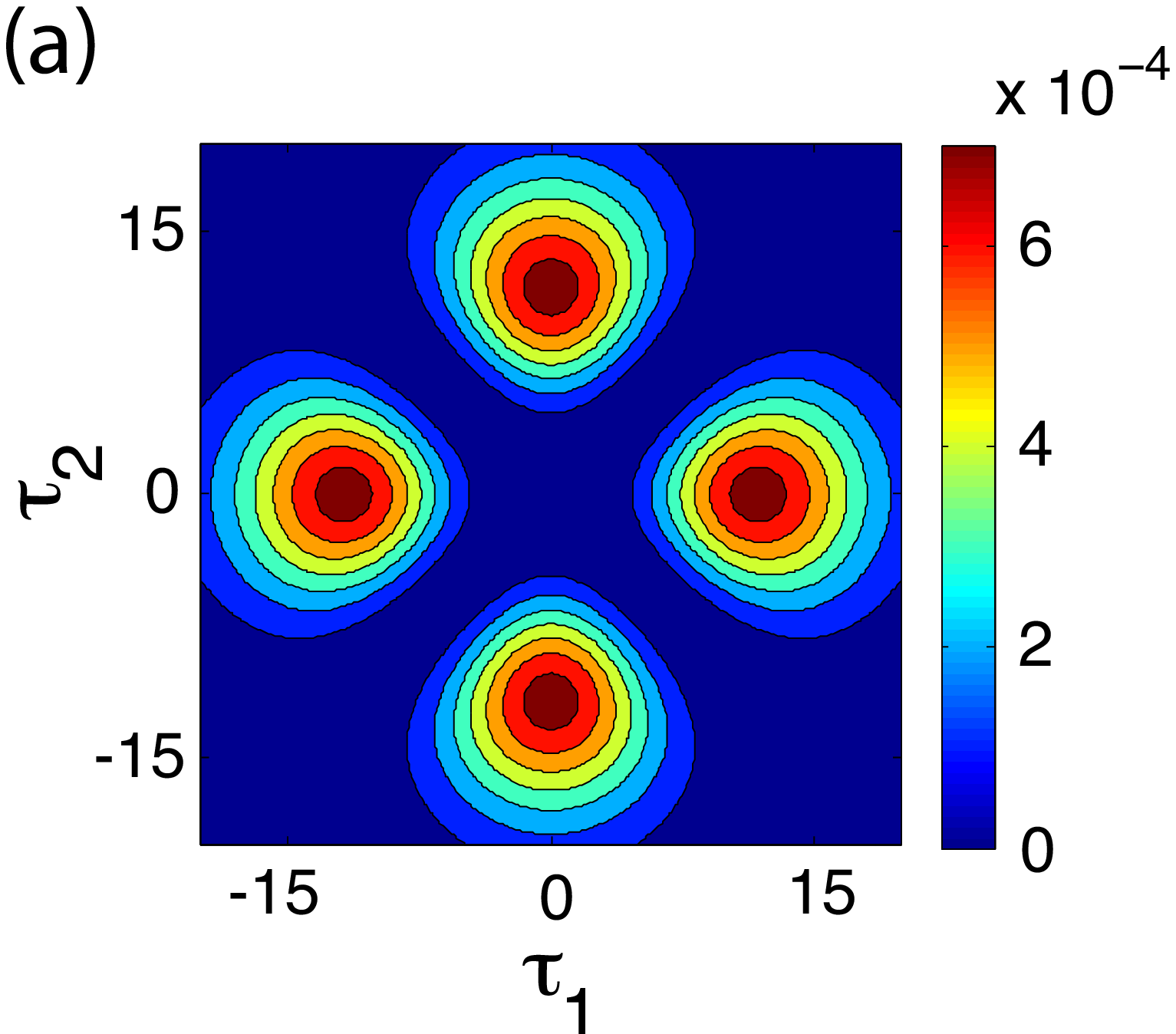}
        \includegraphics[width=4.2cm]{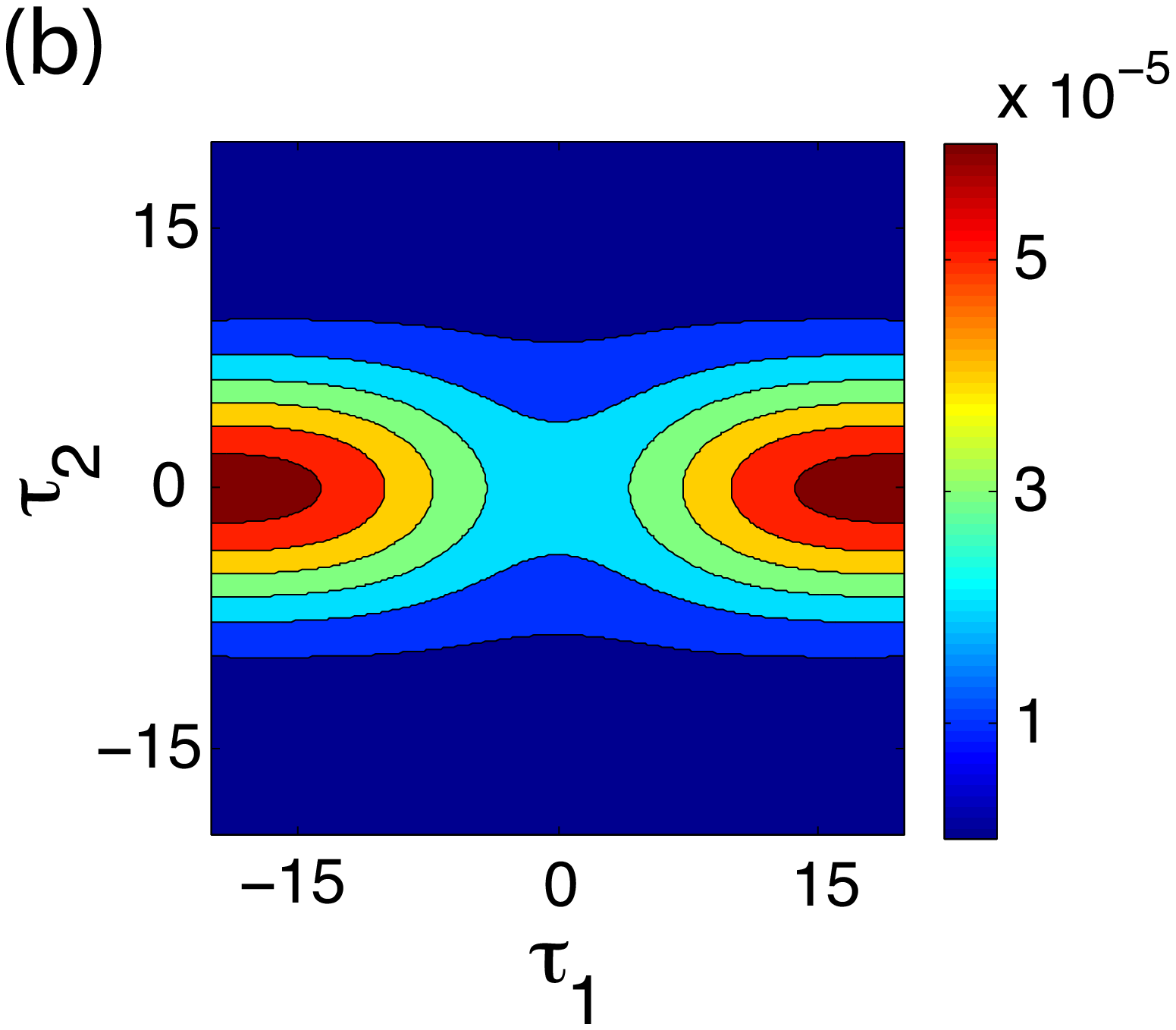}
        \includegraphics[width=4.2cm]{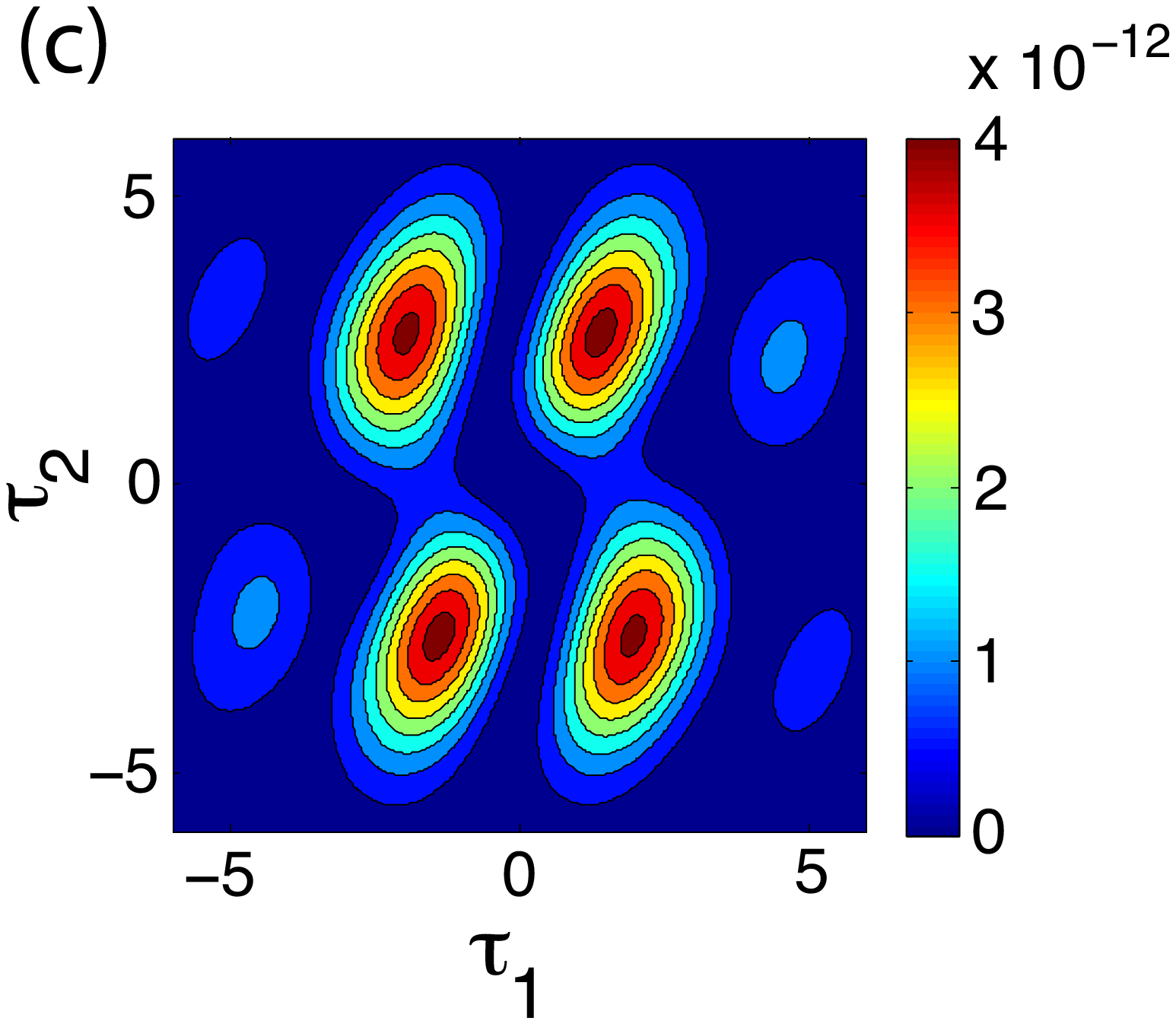}
        \includegraphics[width=4.2cm]{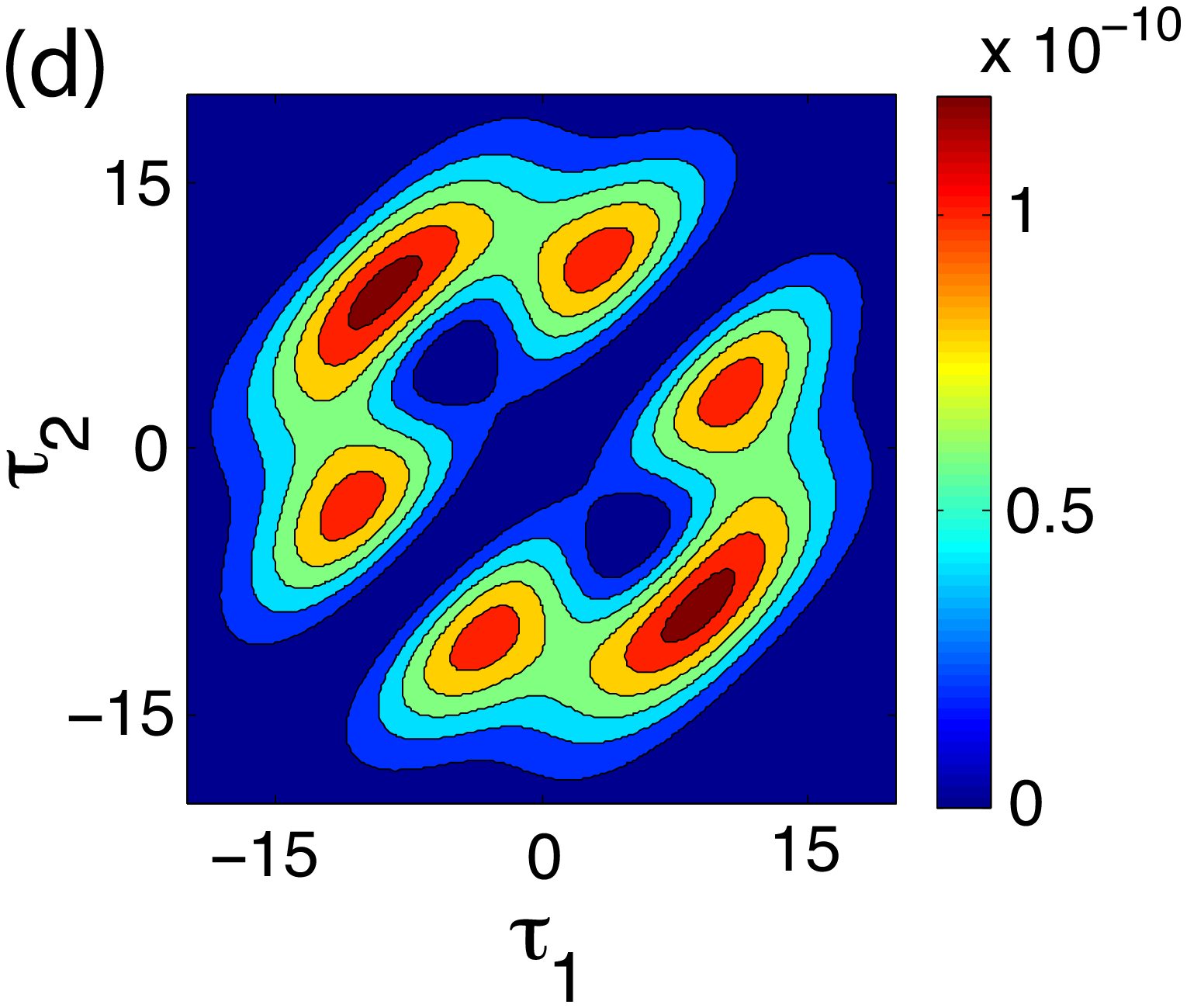}
	\caption{Coincidence rate landscape for three different sets of
		$\Omega\equiv(\alpha_1,\beta_1,\alpha_2,\beta_2,\alpha_3,\beta_3,\gamma_1,\gamma_2)$: 
	(a)~$\Omega=(0,\pi/2,\pi/2,\pi,0,\pi/2,\pi/2,\pi)$, $\omega_0=0$, $\sigma_0=0.1$,
	$\omega_1=0$, $\sigma_1=0.1$, $\omega_2=0$, $\sigma_2=0.1$, and $\omega_3=0$, $\sigma_3=1$, (b)~$\Omega=(0,\pi/2,0,\pi/2,0,\pi/2,0,0)$, $\omega_0=0$, $\sigma_0=1$, $\omega_1=0$, $\sigma_1=0.1$, $\omega_2=0$, $\sigma_2=0.1$ and $\omega_3=0$, $\sigma_3=0.01$, (c)~$\Omega=(0,\pi/2,\pi/2,2\cos^{-1}(1/\sqrt{3}),0,\pi/2,0,0)$, $\omega_0=0$, $\sigma_0=0.5$, $\omega_1=3$, $\sigma_1=0.2$, $\omega_2=2$, $\sigma_2=0.2$ and $\omega_3=1$, $\sigma_3=0.2$, and (d)~same $\Omega$ configuration as (a) but with $\omega_0=0.1$, $\sigma_0=0.1$, $\omega_1=0.95$, $\sigma_1=0.11$, $\omega_2=0$, $\sigma_2=0.1$ and $\sigma_3=0$, $\omega_3=0.99$.
	}
    \label{fig:P111}
\end{figure}

The coincidence rate for output in ports $1,2,3$ 
is 
\begin{equation*}
	P_{111}(\Omega)
		={}^S
			\braket{111|R^\dagger(\Omega)\Pi_1\otimes\Pi_1\otimes\Pi_1
				R(\Omega)|111}^S,
\end{equation*}
which is a function of the two interphoton delay times~$\tau_1$ and~$\tau_2$
and plotted in Fig.~\ref{fig:P111} for various values of~$\Omega$
and choices of the three detector spectral terms
$\omega_i$ and $\sigma_i$, $i=1, 2,$ and $3$.
This rate~$P_{111}(\Omega)$
is a sum of six terms, one for each possible permutation of the three frequencies $(\omega_i,\omega_j,\omega_k)$ of the photons in
output channels $(1,2,3)$ respectively.
Each term is in turn the product of a sum of $D$ functions similar to (\ref{eq:weightedsum}) containing information
on the interferometer through the parameters $\Omega$,
and a detector spectral term as in Eq.~(\ref{eq:gaussian}).

The condition for zero coincidence dip at zero time delays is
\begin{align}
\Lambda_1\Lambda_2\Lambda_3\left|\text{Per}(R(\Omega))\right|^2=0 \ .
\label{permanentcondition}
\end{align}
This relationship shows that a zero coincidence can only be obtained as a result of one of the following two conditions: a zero spectral mismatch or a zero permanent. We emphasize that this second condition is linked to complete indistinguishability of photons at zero time delays. In the more general case of nonzero time delays, which forces partial distinguishability between photons, this requirement will depend on a superposition of immanants.

According to observation 1, interferometer settings that make $D^{(3,0)}_{(111)1;(111)1}(\Omega)$ vanish also make $\text{Per}(R(\Omega))$ vanish from $P_{111}(\Omega)$, thereby resulting in a dip.
 In Figs.~\ref{fig:P111}(a) and (b), we choose
$\Omega$ such that $D^{(3,0)}=0$ and $D^{(3,0)}= -1/4\sqrt{2}$, respectively.
As expected, only Fig.~\ref{fig:P111}(a) has vanishing dip at the origin of the plot
corresponding to two zero time delays. In Fig.~\ref{fig:P111}(b), the dip is not complete for zero time delays. Rather, the coincidence rate is a superposition of immanants at nonzero time delays but equal to the nonzero permanent at zero time delays.
We also provide two examples of three-photon coincidence rates to demonstrate the complexity of dips that can arise when imperfect sources and detectors are used.
In Figs.~\ref{fig:P111}(c) and (d), $D^{(3,0)}=0$,
but detector terms are asymmetric:
contributions from the immanants and determinant interfere,
and we obtain dips at nonzero time delays.
Although here we have focused on the rate of obtaining threefold coincidences
corresponding to detecting one photon at each output port,
we can easily calculate other coincidence rates such as measuring~$\nu_1$ photons exiting the
first port, $\nu_2$ from the second, and~$\nu_3$ from the third port
simply by replacing $\Pi_1\otimes\Pi_1\otimes\Pi_1$
in the calculation of the rate~$P_{111}(\Omega)$
by $\Pi_{\nu_1}\otimes\Pi_{\nu_2}\otimes\Pi_{\nu_3}$.

In summary, we have used SU(3) group theory to calculate the photon-coincidence rates at the 
output of a three-channel interferometer given single photons entering each of the three input
ports. Our analysis of coincidence rates of three-photon coincidences as a function of delay times provides a background coincidence rate against which the depths of coincidence dips can be gauged and shows that a rich array of coincidence dips exists from which immanants of the unitary matrix can be inferred.
Our technique provides a powerful calculational tool for modeling and interpreting 
output data from realistic experiments that are on the cusp of taking the two-photon
two-channel Hong-Ou-Mandel dip to a new and exciting regime of single photons 
entering multichannel interferometers.
Although brute-force calculations can be used to model the outputs from such interferometers,
our methods are much more powerful than typical theoretical quantum optics techniques
for modeling photonic quantum interferometry,
both in terms of making the model tractable and also in making the resultant dips understandable.
We have provided a novel and clear connection between immanants and specific Wigner $D$ functions for SU(3) through the use of Young diagrams. This connection is central to understanding the coincidence rate landscapes as depicted in Fig.~\ref{fig:P111}.
This connection is especially important in light of the potential application of higher-order 
photon-coincidence dips to quantum computation such as for the boson sampling  
problem~\cite{AA11}.

We thank DKL~Oi, and T~Welsh for helpful discussions.
HdG is supported by NSERC and BCS by NSERC, CIFAR and AITF.
YYG acknowledges support from an A*STAR NSS Fellowship.

\bibliography{qHOM_proof}
\bibliographystyle{apsrev4-1.bst}
\end{document}